\documentclass{article}

\usepackage{PRIMEarxiv}

\usepackage[utf8]{inputenc} 
\usepackage[T1]{fontenc}    
\usepackage{hyperref}       
\usepackage{url}            
\usepackage{booktabs}       
\usepackage{amsfonts}       
\usepackage{nicefrac}       
\usepackage{microtype}      
\usepackage{lipsum}
\usepackage{fancyhdr}       
\usepackage{graphicx}       
\graphicspath{{media/}}     
\usepackage{subcaption}
\usepackage{natbib}
\usepackage{amsmath}
\usepackage{algorithmic}
\usepackage{algorithm}
\newcommand{\vect}[1]{\mathbf{#1}}

\title{Diffusion prior as a direct regularization \\ term for FWI}

\author{
  Yuke Xie, Hervé Chauris, Nicolas Desassis \\
  Mines Paris, PSL Research University \\
  77300 Fontainebleau, France\\
}

\begin{document}
\maketitle

\begin{abstract}
Diffusion models have recently shown promise as powerful generative priors for inverse problems. However, conventional applications require solving the full reverse diffusion process and operating on noisy intermediate states, which poses challenges for physics-constrained computational seismic imaging. In particular, such instability is pronounced in non-linear solvers like those used in Full Waveform Inversion (FWI), where wave propagation through noisy velocity fields can lead to numerical artifacts and poor inversion quality. In this work, we propose a simple yet effective framework that directly integrates a pretrained Denoising Diffusion Probabilistic Model (DDPM) as a score-based generative diffusion prior into FWI through a score rematching strategy. Unlike traditional diffusion approaches, our method avoids the reverse diffusion sampling and needs fewer iterations. We operate the image inversion entirely in the clean image space, eliminating the need to operate through noisy velocity models. The generative diffusion prior can be introduced as a simple regularization term in the standard FWI update rule, requiring minimal modification to existing FWI pipelines. This promotes stable wave propagation and can improve convergence behavior and inversion quality. Numerical experiments suggest that the proposed method offers enhanced fidelity and robustness compared to conventional and GAN-based FWI approaches, while remaining practical and computationally efficient for seismic imaging and other inverse problem tasks.
\end{abstract}


\section{Introduction}
Full Waveform Inversion (FWI) is a powerful technique for reconstructing high-resolution subsurface models by minimizing the discrepancy between observed and simulated seismic wavefields \citep{plessix2006review, virieux2009overview, chauris2019fwi}. Despite its potential, the non-linear inverse problem like FWI is highly sensitive to noise and the choice of prior information, making it prone to convergence issues \citep{calvetti2018inverse}, cycle skipping, and poor reconstructions in ill-posed settings. Conventional regularization techniques, such as Tikhonov \citep{golub1999tikhonov}, total variation \citep{strong2003edge, esser2018total}, and sparsity-promoting priors \citep{zhu2017sparse}, help mitigate these challenges but often lack adaptability to complex geological structures.

Recent advances in deep learning \citep{lecun2015deep}, especially deep generative models, such as Generative Adversarial Networks (GAN), which can learn mapping from simple distributions to complex multivariable distributions \citep{goodfellow2020generative}, have demonstrated considerable efficacy in diverse applications in generating realistic images such as fake faces. Denoising Diffusion Probabilistic Model (DDPM) \citep{ho2020denoising} has demonstrated its effectiveness as learned priors for image generation from Gaussian noise. Diffusion models iteratively transform Gaussian noise into structured images by learning the data distribution through a Markovian denoising process. 

In FWI, generative diffusion priors have been introduced to improve inversion stability by constraining solutions to the learned manifold of realistic subsurface models. Deep generative models such as GAN can serve as powerful priors for capturing the complex nature of geophysical parameters \citep{bhavsar_stable_2024, garayt2025two} and serve as the prior distribution for FWI \citep{mosser2020stochastic, fang2020deep, xie2024stochastic}. However, the training of a GAN is not stable  \citep{weng2019wgan}, and it has a strong constraint as priors for solving inverse problems, which provides less variability as a regularization tool in FWI \citep{fang2020deep, xie2024stochastic}.

Notably, prior works have explored applying DDPM for posterior sampling in inverse problems \citep{chung_improving_2024, chung_diffusion_2024}. In seismic imaging, solving the reverse diffusion process to sample plausible models that fit seismic data constraints enhances the FWI imaging quality \citep{wang2023prior, wang2024controllable, shi_generative_2024}. However, these methods require operations through intermediate noisy states of the reverse diffusion process. Injecting DDPM-style noise into the velocity model during wave propagation may lead to non-physical scattering. The added noise introduces artificial perturbations that deviate from subsurface physics, potentially generating false reflectors and leading to unstable inversion. Moreover, high-frequency noise violates the smoothness assumptions required by seismic solvers, which may cause numerical dispersion and instability. Since seismic wave propagation relies on smooth velocity fields for stable finite-difference computations, such noise may lead to the solver being unstable. \cite{graikos2022diffusion} uses an independently trained DDPM model as prior and gives the possibility to turn diffusion models into a direct regularizer for FWI, thereby allowing a range of potential applications in adapting models to more complex constraints such as non-linear equations. 

In this work, we propose an approach that leverages a pretrained DDPM denoiser as a direct regularization term to guide FWI, without requiring explicit sampling from the diffusion process \cite{graikos2022diffusion}. Instead of operating in the noisy latent space, we propose to perform inversion directly in the smooth image space at $t_0$ state, integrating learned priors while avoiding operations on noisy states. The DDPM model plays the role of prior information at each timestep of the FWI iterations. This enables a more stable and computationally efficient inversion process, making it well-suited for both linear and nonlinear inverse problems. Our contributions can be summarized as follows:
\renewcommand{\theenumi}{\roman{enumi}}%
\begin{enumerate}
    \item We propose a direct integration of pretrained generative diffusion priors into FWI, eliminating the need for reverse diffusion sampling.
    \item We demonstrate that our approach allows stable inversion by avoiding noisy intermediate states while preserving the benefits of generative diffusion models.
    \item We validate our method on synthetic seismic data, showing improved convergence and robustness compared to conventional FWI approaches.
\end{enumerate}
The remainder of this paper is organized as follows. In Section 2, we review the theoretical background of DDPMs and their application in inverse problems. Section 3 details our proposed method. Sections 4, 5, and 6 present numerical experiments, and Section 7 concludes with discussions on future directions.

\section{Theory}
In this work, our proposed improved FWI method uses direct diffusion-prior, the prior is an independently trained denoising diffusion generative model (DDPM). And our method uses the pretrained generative denoising diffusion denoiser to serve as a plug-and-play regularization method on conventional FWI updates with minimal modification. This section will give an introduction to the DDPM model and the usage of the DDPM denoiser as a direct prior.

\subsection{Generative Diffusion Models}
\subsubsection{Diffusion process}
Denoising Diffusion Generative Model (DDPM) known as one of the most studied Generative Diffusion Models use the approximate posterior $q(\vect{x}_{1:T} |\vect{x}_0)$, called the forward process or diffusion process, is fixed to a Markov chain that gradually adds Gaussian noise to the data according to a variance schedule $\beta = \{ \beta_1,...,\beta_T \}$, using the notation $\alpha_t:= 1-\beta_t$ and $\bar\alpha_t:= \prod_{s=1}^t \alpha_s$, we define
\begin{equation}
    q(\vect{x}_{1:T} | \vect{x}_0) := \prod_{t=1}^T q(\vect{x}_t | \vect{x}_{t-1} ), \qquad 
\end{equation}
\begin{equation}
    q(\vect{x}_t|\vect{x}_{t-1}) := \mathcal{N}(\vect{x}_t;\sqrt{1-\beta_t}\vect{x}_{t-1},\beta_t \mathbf{I}) \label{eq:forwardprocess}
\end{equation}
where $q(\vect{x}_t | \vect{x}_{t-1} )$ is the forward transition.

The forward diffusion process adds noise step-by-step:
\begin{equation}
    x_t = \sqrt{\alpha_t} x_{t-1} + \sqrt{1-\alpha_t} \epsilon_t
\end{equation}
where \(\epsilon_t \sim \mathcal{N}(0, \mathbf{I})\) is standard Gaussian noise.

For multiple steps, this schedule variance rule enables the training by efficiently sampling from the conditional distribution
\begin{equation}
    q(\vect{x}_t|\vect{x}_{0}) := \mathcal{N}(\vect{x}_t;\sqrt{\bar\alpha_t}\vect{x}_{0}, (1-\bar\alpha_t) \mathbf{I})
\end{equation}

To sample a single point in this distribution, we may sample \( \boldsymbol{\epsilon} \sim \mathcal{N}(\mathbf{0}, \mathbf{I}) \) as part of the diffusion process:
\begin{equation}\label{eq:shortcut}
\mathbf{x}_t = \sqrt{\bar{\alpha}_t} \mathbf{x}_0 + \sqrt{1 - \bar{\alpha}_t} \boldsymbol{\epsilon}
\end{equation}
where $ \mathbf{x}_t $ is a noisy version of the image, and $\boldsymbol{\epsilon} $ is the noise added to it.

\subsubsection{Reverse diffusion process}
We can reverse the above process and sample from $q(\mathbf{x}_{t-1} \vert \mathbf{x}_t)$ by recreating a true sample from a Gaussian noise $\mathbf{x}_T \sim \mathcal{N}(\mathbf{0}, \mathbf{I})$, and the reverse conditional probability is tractable when conditioned on $\vect{x}_0$:

\begin{equation}\label{eq:reverse_q}
    q(\mathbf{x}_{t-1} | \mathbf{x}_t, \mathbf{x}_0) = \mathcal{N}(\mathbf{x}_{t-1}; \tilde{\mu}_t(\mathbf{x}_t, \mathbf{x}_0), \tilde{\beta}_t \mathbf{I}),
\end{equation}

we can represent $\mathbf{x}_0 = \frac{1}{\sqrt{\bar{\alpha}_t}}(\mathbf{x}_t - \sqrt{1 - \bar{\alpha}_t}\boldsymbol{\epsilon}_t)$ from Equation \ref{eq:shortcut}, and with reparameterization proposed by \cite{ho2020denoising}, the mean is represented as

\begin{equation}\label{eq:mean_prediction}
\tilde{\boldsymbol{\mu}}_t = \frac{1}{\sqrt{\alpha_t}} \Big( \mathbf{x}_t - \frac{1 - \alpha_t}{\sqrt{1 - \bar{\alpha}_t}} \boldsymbol{\epsilon}_t \Big).
\end{equation}

\subsubsection{Learned diffusion process}
The key of the denoising diffusion model is to learn the reverse process by learning a model $p_{\theta}(\mathbf{x}_{t-1} | \mathbf{x}_t)$ to approximate the conditional probabilities in Equation \ref{eq:reverse_q} at each time step. The joint distribution $p_\theta(\vect{x}_{T:0})$ will represent the reverse process after $\theta$ is learned. It is defined as a Markov chain with learned Gaussian transitions starting at $p(\vect{x}_T)=\mathcal{N}(\vect{x}_T; \mathbf{0}, \mathbf{I})$:
\begin{equation}
    p_\theta(\vect{x}_{T:0}) := p(\vect{x}_T)\prod_{t=1}^T p_\theta(\vect{x}_{t-1}|\vect{x}_t),
\end{equation}
\begin{equation}
      p_\theta(\vect{x}_{t-1}|\vect{x}_t) := \mathcal{N}(\vect{x}_{t-1}; \mathbf{\mu}_\theta(\vect{x}_t, t), \mathbf{\Sigma}_\theta(\vect{x}_t, t))
\end{equation}

The loss function is defined as the KL divergence \citep{kullback1951information} between the true reverse diffusion process and the learned reverse diffusion process 
\begin{equation}\label{eq:loss}
    L :=\mathbb{E}_q \left[ \text{KL}\left(q(\vect{x}_{T:1}|\vect{x}_0) \parallel p_\theta(\vect{x}_{T:0})\right) \right],
\end{equation}
which we can evaluate separately at each timestep,
\begin{equation}\label{eq:loss_t}
    L_{t-1} = \mathbb{E}_q \left[ \text{KL}\left(q(\mathbf{x}_{t-1}|\mathbf{x}_t, \mathbf{x}_0) \parallel p_\theta(\mathbf{x}_{t-1}|\mathbf{x}_t)\right) \right].
\end{equation}
Due to the property of the Gaussian distribution, Equation \ref{eq:loss_t} can be rewritten as
\begin{equation}
\begin{split}
    L_{t-1} &= \mathbb{E}_q \left[ \text{KL}\left(q(\mathbf{x}_{t-1}|\mathbf{x}_t, \mathbf{x}_0) \parallel p_\theta(\mathbf{x}_{t-1}|\mathbf{x}_t)\right) \right]\\ 
    &= \mathbb{E}_{\mathbf{x}_0, \boldsymbol{\epsilon}} \Big[\frac{1}{2 \sigma_t^2} \| {\tilde{\boldsymbol{\mu}}_t(\mathbf{x}_t, \mathbf{x}_0)} - {\boldsymbol{\mu}_\theta(\mathbf{x}_t, t)} \|^2 \Big] 
\end{split}
\end{equation}
Using the reparameterization in Equation \ref{eq:mean_prediction}, 
\begin{equation}
    L_{t-1} = w(t)\ \mathbb{E}_{\vect{x}_0, \vect{\epsilon}}\Big[ { \left\| \mathbf{\epsilon} - \mathbf{\epsilon}_\theta(\sqrt{\bar\alpha_t} \vect{x}_0 + \sqrt{1-\bar\alpha_t}\mathbf{\epsilon}, t) \right\|^2} \Big]
\end{equation}
where $w(\beta_t) = \frac{\beta_t^2}{2\sigma_t^2 \alpha_t (1-\bar\alpha_t)}$ is a weighting factor on the score function.
To minimize the $\mathcal{l}_2$ norm above, we take a stochastic gradient descent step on $t \sim [1, T]$ and $\epsilon \sim \mathcal{N}(\mathbf{0}, \mathbf{I})$
\begin{equation}\label{eq:gradient}
    \frac{\partial L_{t-1}}{\partial \theta} = 
    \nabla_\theta \left\| \mathbf{\epsilon} - \mathbf{\epsilon}_\theta(\sqrt{\bar\alpha_t} \vect{x}_0 + \sqrt{1-\bar\alpha_t}\mathbf{\epsilon}, t) \right\|^2
\end{equation}
Thus, by training $\mathbf{\epsilon}_\theta$ to minimize the loss in Equation \ref{eq:gradient}, the model implicitly learns a score function $\nabla_\vect{x} \log p_t(\vect{x}_t)$ that guides the reverse diffusion process. This score function can be leveraged for inverse problems, where it acts as a generative prior to constrain reconstructions toward the learned data manifold.

\section{Score-based generative diffusion models as prior}
\subsection{Problem setting}
To solve an inverse problem, to obtain the observation we solve the forward problem
\[
\vect{d}=\mathcal{F}(\vect{m}),
\]
where $\mathcal{F}$ is a linear or non-linear forward operator, and the observation is considered follow the physics $\vect{d}_{\text{obs}}=\mathcal{F}(\vect{m}_{\text{true}})$. In general, we want to find an approximation to the posterior distribution
\[p_\theta(\vect{m}|\vect{d}) \propto p_\theta(\vect{m})p(\vect{d}|\vect{m}),\]
where $p_\theta(\vect{m})$ is a fixed prior distribution. Fixing the observation $\vect{d}$ and introducing an approximate variational distribution $q(\vect{x})$ to approximate the posterior. As the definition of variational Bayesian inference, we minimize the KL divergence
\begin{equation}\label{eq:loss_inference}
\begin{split}
    & KL[q(\vect{m})||p_\theta(\vect{m}|\vect{d})]\\
    =\ & \mathbb{E}_{q(\vect{m})} [\log q(\vect{m}) - \log p(\vect{m}|\vect{d})]\\
    =\ & \mathbb{E}_{q(\vect{m})} [\log q(\vect{m}) - \log p(\vect{m}) - \log p(\vect{m}|\vect{d})] + \vect{c}
\end{split}
\end{equation}
is minimized when $q(\vect{m})$ is closest to the true posterior. By neglating the constant term $\vect{c}$ in Equation \ref{eq:loss_inference}, we call it variational free energy, it is a variational bound on the log-evidence and indirectly approximates the posterior $\log p_\theta(\vect{m}|\vect{d})$. Minimizing KL divergence (or maximizing ELBO) is a way to find the best approximation of the true posterior using a simpler, tractable variational distribution $q(\vect{m})$. We are interested in a general procedure to minimize the variational objective function $F$ with respect to an approximate posterior $q(\vect{m})$ for any differentiable $\log p_\theta(\vect{m}|\vect{d})$ when $p_\theta(\vect{m})$ is a DDPM prior.

\subsection{Denoising diffusion probabilistic models as priors}
Using a pretrained DDPM as prior involves using intermediate latent variables $\vect{h} = \{\vect{x}_T,...,\vect{x}_1\}$, When the prior involves latent variables $\vect{h}$, for any input variational distribution $q(\vect{x})$ following the DDPM process, the joint probability can be rewrite as $q(\vect{h}|\vect{x})q(\vect{x})$, the process $p(\vect{x})$ can be rewrite as $p(\vect{x}, \vect{h})$. 

We can then rewrite Equation \ref{eq:loss_inference} by expanding it to the entire diffusion process:
\begin{equation}\label{eq:loss_inference}
\begin{split}
& \mathbb{E}_{q(\vect{h}|\vect{x})q(\vect{x})} [q(\vect{h}|\vect{x})q(\vect{x})  - \log p_\theta(\vect{x, h}) - \log p(\vect{x}|\vect{d})]\\
= &\mathbb{E}_{q(\vect{h}|\vect{x})q(\vect{x})} [q(\vect{h}|\vect{x})q(\vect{x}) - \log p_\theta] - \mathbb{E}_{q(\vect{x})}[\log p(\vect{x}|\vect{d})]
\end{split}
\end{equation}


As introduced in the previous section, DDPM is trained under reversing the (Gaussian) noising process, so we should rewrite $p(\vect{h}, \vect{x})$ following the process by:
\begin{equation}
\begin{split}
    p_\theta(\vect{h}, \vect{x}_0) &= p_\theta(\vect{x}_T, \vect{x}_{T-1}, ..., \vect{x}_1, \vect{x}_0)\\ &= p(\vect{x}_T)\prod_{t=1}^T p_\theta(\vect{x}_{t-1}|\vect{x}_t)
\end{split}
\end{equation}

For the reverse process, we can rewrite:
\begin{equation}
q(\vect{h} | \vect{x}_0) = q(\vect{x}_T | \vect{x}_0) \prod_{t=1}^{T} q(\vect{x}_{t-1} | \vect{x}_t, \vect{x}_0)
\end{equation}
This describes how we move backward from \( x_T \) to \( x_0 \), using \( x_0 \) as a guiding condition.

If we search for a single-point variational estimation by using
\begin{equation}
    q(\vect{x}) = \delta(\vect{x}-\vect{m})
\end{equation}
where $m$ is the initial guess and can commonly be sampled from the data manifold (training images). So we can sample $\vect{x}_t$, which is the noisy version of $\vect{m}$ at an arbitrary time step $t$ using
\begin{equation}
    \mathbf{x}_t = \sqrt{\bar{\alpha}_t} \vect{m} + \sqrt{1 - \bar{\alpha}_t} \boldsymbol{\epsilon}
\end{equation}

So the first two terms in Equation \ref{eq:loss_inference} can be evaluated over time
\begin{equation}
    \sum_t \text{KL}\left(q(\mathbf{x}_{t-1}|\mathbf{x}_t, \mathbf{m}) \parallel p_\theta(\mathbf{x}_{t-1}|\mathbf{x}_t)\right)
\end{equation}

At each time step, using the parameterization in Equation \ref{eq:mean_prediction}
\begin{equation}
\begin{split}
&\sum_t \ \text{KL}\left(q(\mathbf{x}_{t-1}|\mathbf{x}_t, \mathbf{m}) \parallel p_\theta(\mathbf{x}_{t-1}|\mathbf{x}_t)\right)\\
= &\sum_t \mathbb{E}_{\boldsymbol{\epsilon} \sim \mathcal{N} (\mathbf{0}, \mathbf{I})} \Big[\frac{1}{2 \sigma_t^2} \| {\tilde{\boldsymbol{\mu}}_t\big(\mathbf{x}_t( \mathbf{m}, \boldsymbol{\epsilon})\big)} - {\boldsymbol{\mu}_\theta \big(\mathbf{x}_t( \mathbf{m}, \boldsymbol{\epsilon})\big), t)} \|^2_2 \Big]  \\
= &\sum_t w(\beta_t) \ \mathbb{E}_{\boldsymbol{\epsilon} \sim \mathcal{N} (\mathbf{0}, \mathbf{I})} \left\| \boldsymbol{\epsilon} - \boldsymbol{\epsilon}_\theta\big(\mathbf{x}_t( \mathbf{m}, \boldsymbol{\epsilon}), t\big) \right\|^2_2 \Big]
\end{split}
\end{equation}

where $w(\beta_t)$ is a weighting function that can be negated. So the inference loss function in \ref{eq:loss_inference} simplifies to a score rematching objective

\begin{equation}
\begin{split}
    \sum_t \mathbb{E}_{\boldsymbol{\epsilon} \sim \mathcal{N} (\mathbf{0}, \mathbf{I})} \left\| \boldsymbol{\epsilon} - \boldsymbol{\epsilon}_\theta(\mathbf{x}_t, t) \right\|^2_2 \Big]
    - \mathbb{E}_{q(\vect{m})} \left[\log p(\mathbf{d}| \mathbf{m})\right] ,\\ 
    = \sum_t \underbrace{\mathbb{E}_{\boldsymbol{\epsilon} \sim \mathcal{N} (\mathbf{0}, \mathbf{I})} \left\| \boldsymbol{\epsilon} - \boldsymbol{\epsilon}_\theta(\mathbf{x}_t, t) \right\|^2_2 \Big]}_{\text{Prior term (Score Rematching)}} 
    + \underbrace{\frac{\left\| \mathcal{F}(\vect{m}) - \vect{d}_\text{obs} \right\|^2_2}{2\sigma_\text{noise}}}_{\text{Data Misfit term}} ,\\ 
    \text{where} \quad  
    \mathbf{x}_t(\vect{m}, \boldsymbol{\epsilon}) = \sqrt{\bar{\alpha}_t} \boldsymbol{\vect{m}} + \sqrt{1-\bar{\alpha}_t} \boldsymbol{\epsilon}.
    \label{eq:inferece_update}
\end{split}
\end{equation}

\begin{figure*}
    \centering
    \includegraphics[width=1\linewidth]{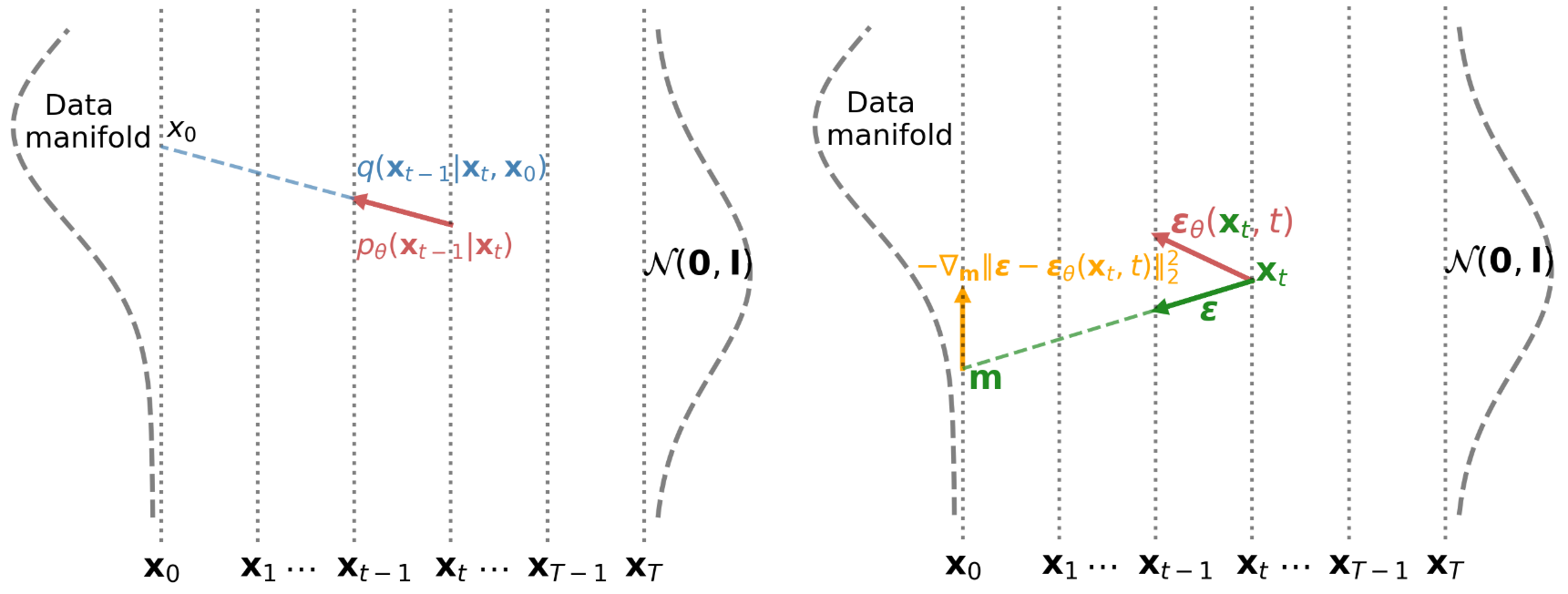}
    \caption{\textbf{a} Diagram: the model is trained to remove noise from given time step $t$. \textbf{b} Using pretrained DDPM to compute the gradient of the clean image $\vect{m}$ towards the data manifold by re-matching the score.}
    \label{fig:diagram_ddpm_prior}
\end{figure*}

Figure \ref{fig:diagram_ddpm_prior}a showcases the diagram that the model is trained to remove noise from the given time step $t$ using a U-Net. \textbf{b} Using pretrained DDPM to compute the gradient of the clean image $\vect{m}$ towards the data manifold by differentiating and minimizing the denoising score function, the gradient is computed to directly update the clean velocity image $\vect{m}$.

\begin{figure*}
    \centering
    \includegraphics[width=0.7\linewidth]{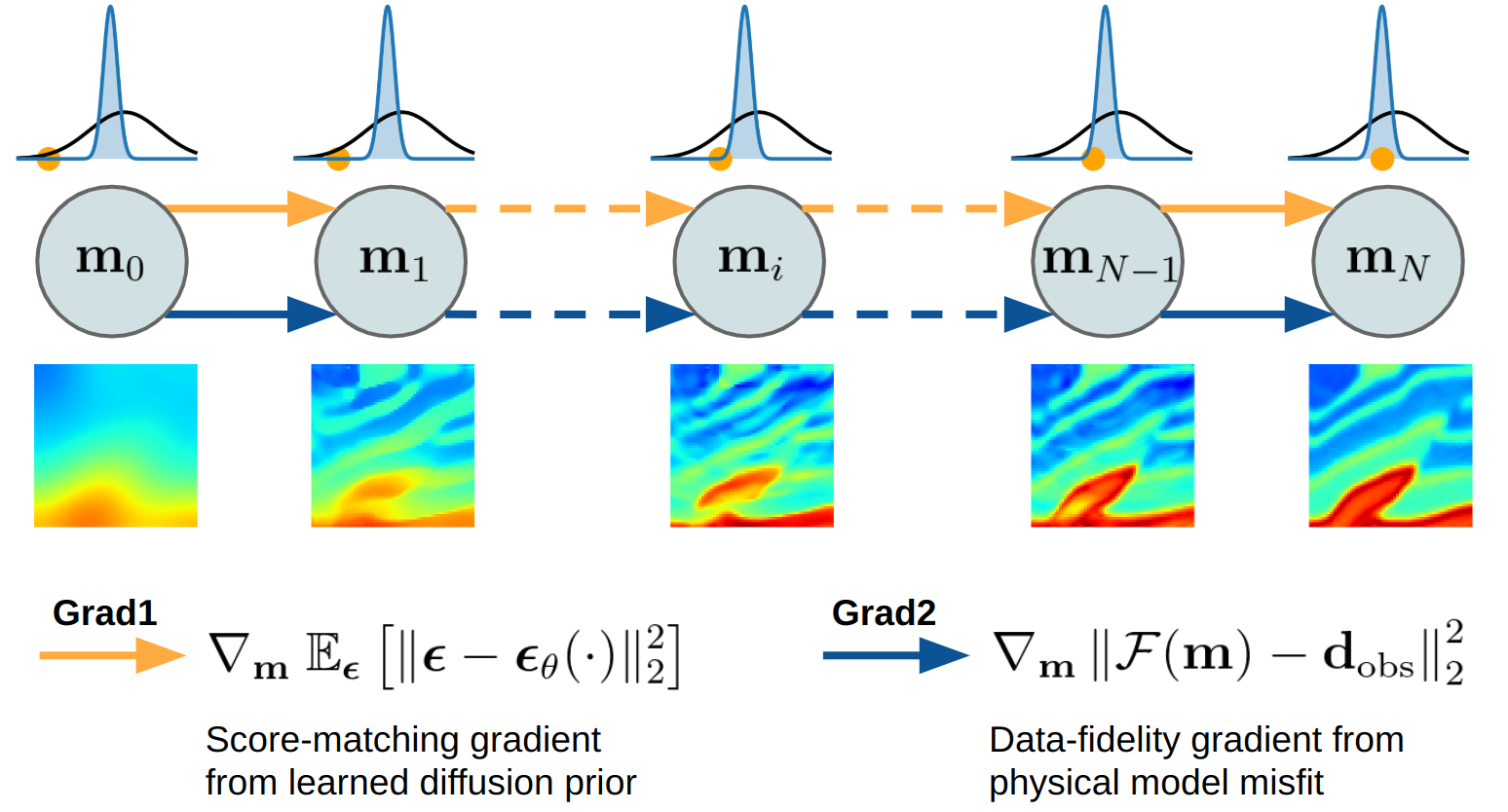}
    \caption{
        A diagram illustrating the optimization process with two gradient components. 
        The orange arrow represents the score-matching gradient from the learned DDPM prior (Grad1), which encourages solutions to lie on the data manifold.
        The blue arrow represents the data misfit gradient (Grad2), driving the solution to match the observed data through the forward model.}
    \label{fig:diagram_algorithm}
\end{figure*}

\begin{algorithm*}
\caption{FWI with a diffusion regularization term}\label{alg:optimization}
\begin{algorithmic}[1]
    \STATE \textbf{Input:} pretrained DDPM denoiser $\boldsymbol{\epsilon}_\theta$, observed data $\boldsymbol{d}$, annealing schedule $t = \{t_{i=0}, t_{i=1}, \dots, t_{i=N}\}$, learning rate $\lambda_t$, weight schedule $w_t$, forward operator $\mathcal{F}$, observation $\vect{d_\text{obs}}$

    \STATE Choose clean initial guess image $\vect{m}_0$.
    \FOR {$i = 0$ \TO $N$}
        \STATE Reverse diffusion time step $t = t_i$
        \STATE Sample batch $\boldsymbol{\epsilon} = \{\boldsymbol{\epsilon}_0, \dots, \boldsymbol{\epsilon}_{n} \} \sim \mathcal{N}(\boldsymbol{0}, \mathbf{I})$
        \STATE  $\mathbf{m} \leftarrow \mathbf{m} - \lambda_t \nabla_{\mathbf{m}} 
        \Big[ \mathbb{E}_{\boldsymbol{\epsilon}} \left[ \left\|\boldsymbol{\epsilon} - \boldsymbol{\epsilon}_{\theta}(\sqrt{\bar{\alpha}_{t}}\mathbf{m} + \sqrt{1-\bar{\alpha}_{t}}\boldsymbol{\epsilon},t)\right\|_2^2\right] - w_t \left\| \mathcal{F}(\vect{m}) - \vect{d}_\text{obs} \right\|^2_2\Big]$
    \ENDFOR
    \RETURN $\vect{m}^*$
\end{algorithmic}
\end{algorithm*}

\section{Training the Denoising Diffusion Probabilistic Model}

\begin{figure}
    \centering
    \includegraphics[width=0.4\linewidth]{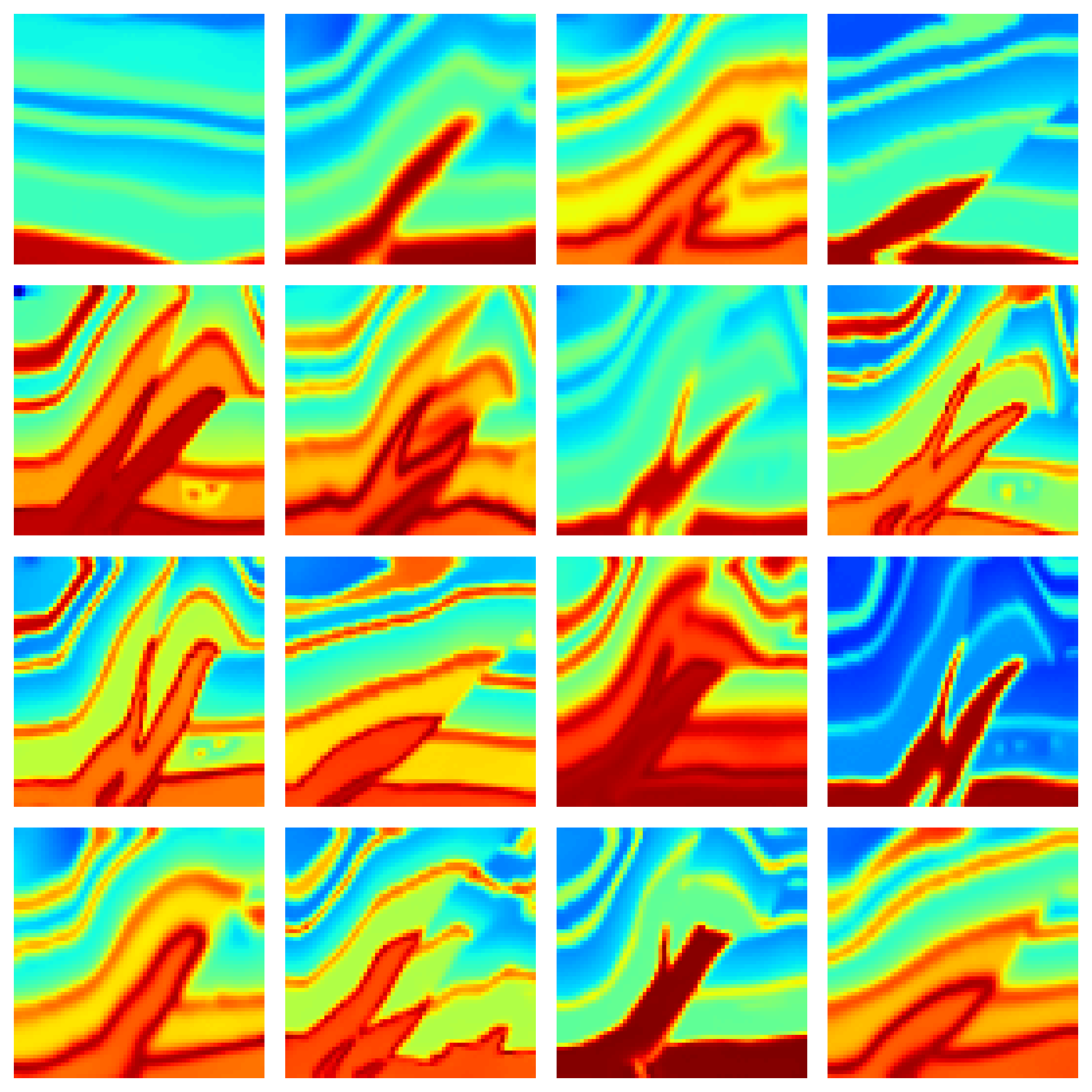}
    \caption{Generated samples using pretrained DDPM, the images represent the data manifold of the prior distribution.}
    \label{fig:generated}
\end{figure}

\begin{figure*}
    \centering
    \includegraphics[width=0.85\linewidth]{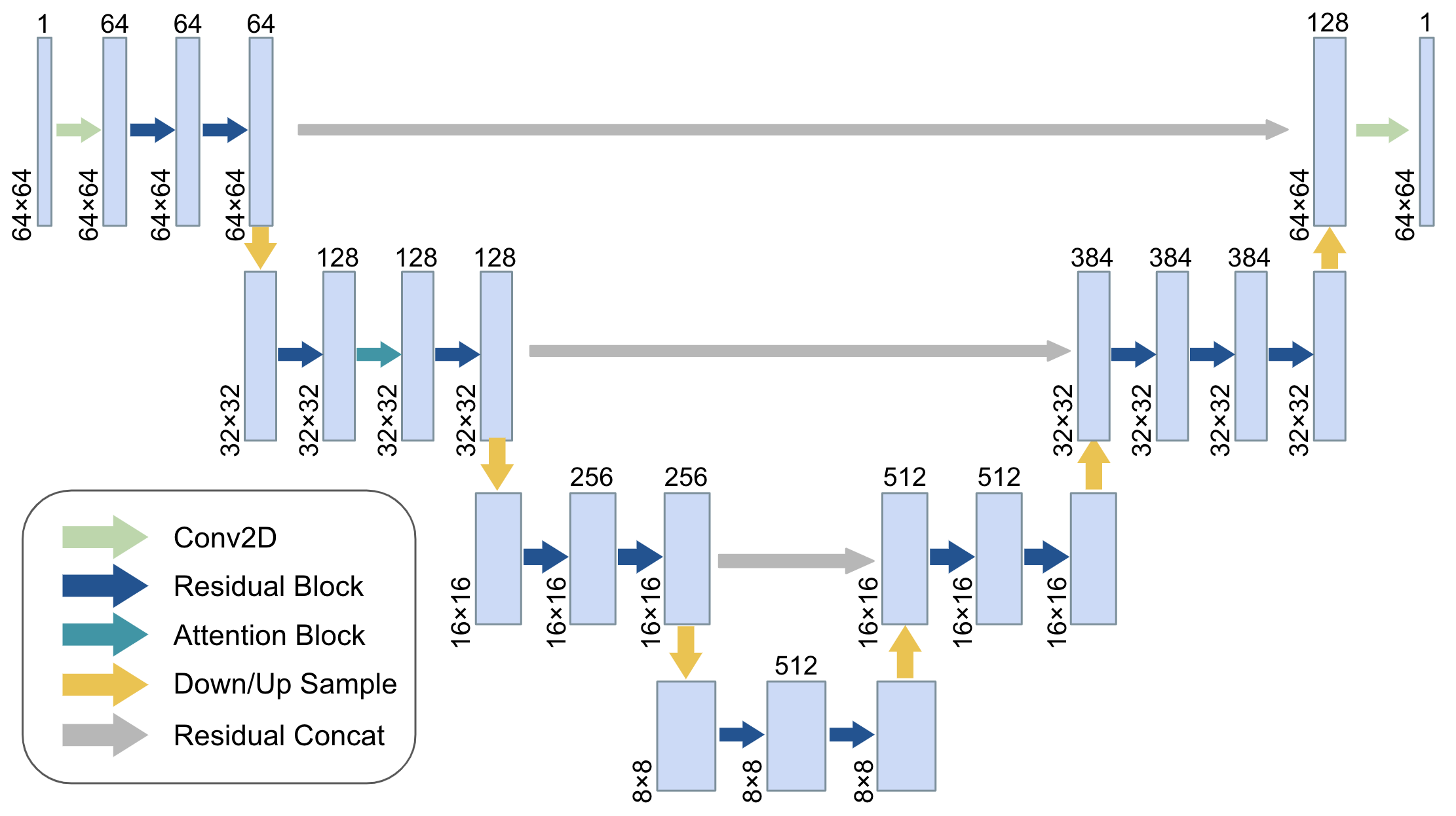}
    \caption{Diagram of the U-Net architecture used for noise prediction in DDPM. The network consists of several down-sampling and up-sampling blocks with attention blocks. The input image is progressively denoised using predicted noise at each timestep of the diffusion process.}
    \label{fig:unet}
\end{figure*}

In this work, we train a DDPM denoiser following the Algorithm \ref{alg:ddpm_training} using a U-Net-based architecture to predict the noise at each timestep. The model is designed to work with grayscale images of size $ 64 \times 64 $ pixels, which represent velocity Vp, and it employs a series of down-sampling and up-sampling blocks, with attention mechanisms incorporated in the deeper layers. 

Specifically, the model consists of 4 down-sampling and 4 up-sampling blocks, with the number of channels increasing progressively at each block according to a factor of 2, starting from 64 channels in the first convolutional layer. A diagram of the U-Net architecture used for noise prediction is shown in Figure \ref{fig:unet}. The network also incorporates Group Normalization with 8 groups to normalize the activations across the layers. Attention mechanisms are applied in the third and fourth down-sampling blocks. The U-Net receives both the image and a time embedding vector as input. The time embedding is generated using a dedicated Time Embedding module that processes the timestep $t$. During training, the model learns to predict the noise added to the image at each timestep in the diffusion process. 

The batch size was set to 32 for training, and the model was trained for 80 epochs. The total number of timesteps in the diffusion process was set to $T=1000$. The beta schedule, which controls the noise variance at each timestep, followed a linear progression starting from $ \beta_{\text{start}} = 1 \times 10^{-4} $ and ending at $ \beta_{\text{end}} = 0.02 $. The learning rate for the Adam optimizer was set to $ 2 \times 10^{-4} $, with the default Adam parameters $ \beta_1 = 0.9 $ and $ \beta_2 = 0.999 $. We also used the clipping range of $[-1.0, 1.0]$ to ensure that pixel values remained within a valid range during training. 

The dataset used for training was a modified version \citep{xie2024stochastic} derived from the Overthrust dataset \citep{aminzadeh19963overthrust}, consisting of 40,000 images. Training was conducted on a single external Nvidia GTX 1080 eGPU with 8GB of RAM. We trained the model for a total of 120 epochs, with each epoch taking $1694 \pm 15.2$ seconds (approximately 28 minutes). The DDPM was trained to predict and remove the added noise at each diffusion timestep via a score-matching objective. Figure \ref{fig:generated} shows generated samples from the trained model, which reflect the learned data manifold and illustrate the ability of the model to capture meaningful geological structural prior from the training data distribution.

The U-Net architecture serves as the backbone for the noise prediction task in the DDPM. In the normal image generation process, given a noisy image and a corresponding timestep, the network outputs a predicted noise map, which is then used to reverse the noise process. This iterative process continues for each timestep, progressively denoising the image until the final clean image is recovered. 

\begin{algorithm*}
\caption{Training a Denoising Diffusion Probabilistic Model (DDPM) denoiser}\label{alg:ddpm_training}
\begin{algorithmic}[1]
    \STATE \textbf{Input:} Data $\mathcal{X} = \{ \mathbf{x}_1, \mathbf{x}_2, \dots, \mathbf{x}_N \}$, noise schedule $\beta_1, \dots, \beta_T$, U-Net $\epsilon_\theta$, learning rate $\eta$
    \STATE Initialize model parameters $\theta$.
    \FOR {epoch = 1 \TO EPOCHS}
        \FOR {batch = 1 \TO BATCHES}
            \STATE Sample a batch of images $\{ \mathbf{x}_i \}_{i=1}^{B}$ from data $\mathcal{X}$.
            \STATE Sample random time step $t \sim \text{Uniform}(1, T)$.
            \STATE Sample noise $\mathbf{\epsilon} \sim \mathcal{N}(\mathbf{0}, \mathbf{I})$.
            \STATE Generate noisy image $\mathbf{x}_t = \sqrt{\bar{\alpha}_t} \mathbf{x}_0 + \sqrt{1-\bar{\alpha}_t} \mathbf{\epsilon}$ where $\bar{\alpha}_t = \prod_{s=1}^t \alpha_s$.
            \STATE Compute the loss function:
            \[
                L_{\text{ddpm}}(\theta) = \mathbb{E}_{t, \mathbf{x}_0, \mathbf{\epsilon}} \left[ \left\| \mathbf{\epsilon} - \epsilon_\theta(\mathbf{x}_t, t) \right\|^2_2 \right]
            \]
            \STATE Update parameters $\theta \leftarrow \theta - \eta \nabla_\theta L_{\text{ddpm}}(\theta)$.
        \ENDFOR
    \ENDFOR
    \RETURN $\epsilon_\theta$
\end{algorithmic}
\end{algorithm*}

\section{Results}
Following the training of the U-Net, we employed the mentioned methods using the DDPM denoiser to regularize FWI.  We compute the gradient of the score rematching diffusion prior term in the first term of \ref{eq:inferece_update} using TensorFlow automatic differentiation function, and the data misfit term in second term of \ref{eq:inferece_update} is computed by solving wave equation and adjoint wave equation, the numerical computations are executed utilizing GPU acceleration by Cupy \citep{cupy_learningsys2017}. PDEs are solved numerically using a second-order finite difference scheme. The true velocity model is cropped from the original Overthrust dataset, the computational grid comprises 64-depth ($z$) by 64-width ($x$) grids. 

In our numerical experiments, we solve the acoustic wave equation using a frequency cap of 25Hz, processing 600 iterations in the time domain with a discrete interval of 1ms. The spatial grid points, set at $9.5 \text{m}$ intervals for both 
$dz$ and $dx$, and with grid points spaced at intervals of 8.71 meters in both the z and x directions. Receivers are distributed across every grid point on the surface, with different source configurations on the surface of the model. We assume that a Gaussian distribution can represent the noise $\sigma_{\text{noise}}$ of the seismic data in the likelihood function, where the likelihood function influences the weighting or strength of belief assigned to the prior term. And we use a scheduled weighting function during FWI optimization such that we negate the $\sigma_{\text{noise}}$ term.

\begin{figure*}
    \centering
    \includegraphics[width=1\textwidth]{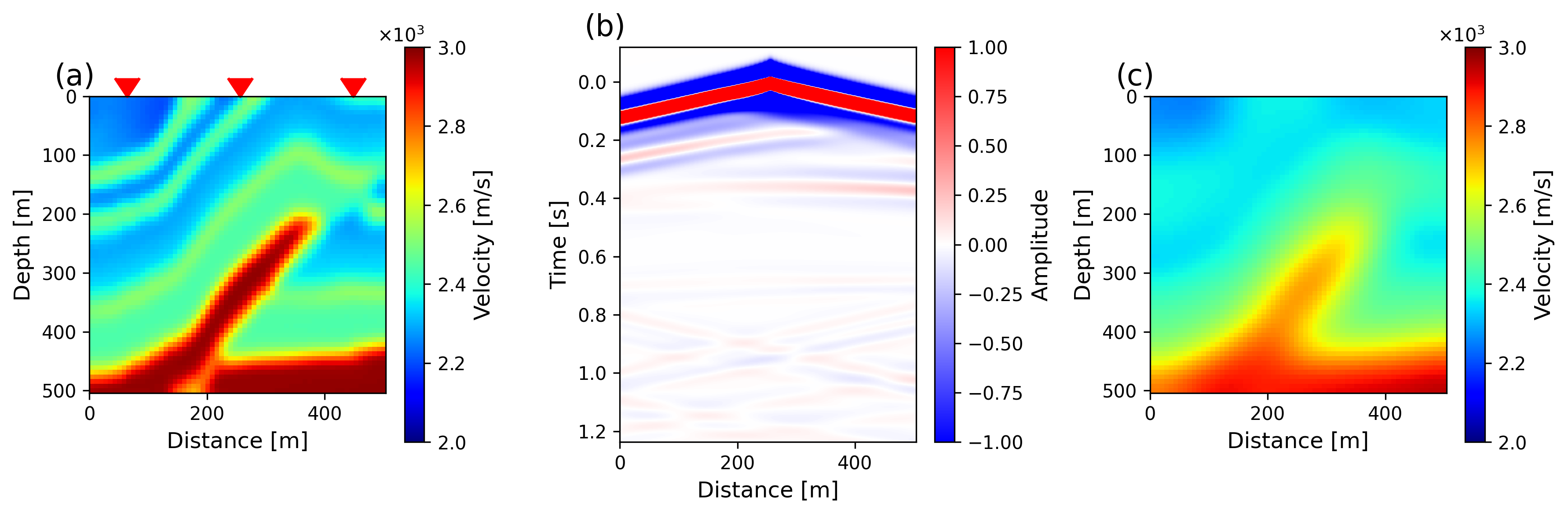}
    \caption{Visualization of the velocity model depicting the subsurface structure, with receivers distributed across every grid point along the surface of the model. \textbf{a} True velocity model and seismic source configuration. \textbf{b} One example of a simulated seismic shot gather. \textbf{c} Initial velocity model used as the starting point for both conventional FWI and diffusion-prior FWI.}
    \label{fig:forward}
\end{figure*}

\begin{figure}
    \centering
    \includegraphics[width=0.5\textwidth]{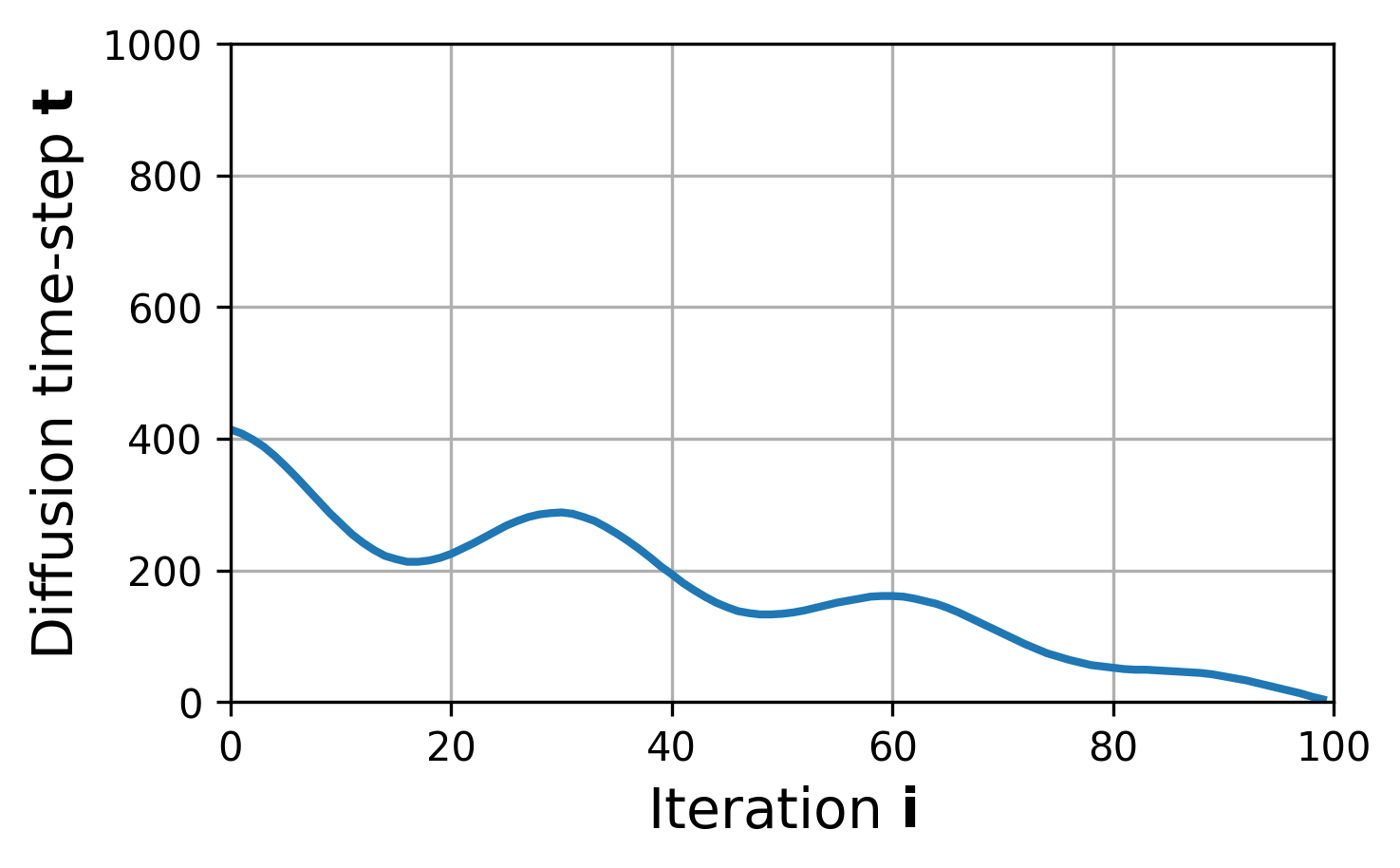}
    \caption{Predefined diffusion time-step $t$ annealing schedule for evaluating the score-matching function across inversion iterations $i = 0$ to $i = 99$, constructed by superimposing a cosine perturbation on a linear decay function.}
    \label{fig:t_schedule}
\end{figure}

Figure~\ref{fig:forward}\textbf{a} illustrates the true subsurface velocity model and the seismic source configuration. The corresponding synthetic shot gather is shown in Figure~\ref{fig:forward}\textbf{b}. The same initial velocity model used as the starting point for both conventional and diffusion-prior FWI is presented in Figure~\ref{fig:forward}\textbf{c}.

Conventional FWI is performed by computing the gradient of the data misfit using the adjoint-state method and applying this gradient to update the velocity model. For the proposed diffusion-prior FWI, we evaluate the score-rematching loss at each iteration using a pre-defined diffusion time-step schedule, as illustrated in Figure~\ref{fig:t_schedule}. The diffusion time-step $t$ is annealed from large to small values throughout 100 iterations using a linearly decaying function modulated by a cosine perturbation inspired by \citep{graikos2022diffusion}.

To balance the influence of the diffusion-prior with the data misfit term, we normalize the gradient of the FWI relative to the diffusion-prior gradient. A linear weighting schedule is applied to the diffusion-prior term, increasing from 0.1 to 0.2 across iterations. This strategy ensures that the early stages of inversion are primarily guided by seismic data misfit, while the influence of the learned prior becomes more prominent in later iterations, refining the model towards the learned data manifold.

\begin{figure}[!htp]
    \centering
    \includegraphics[width=0.75\textwidth]{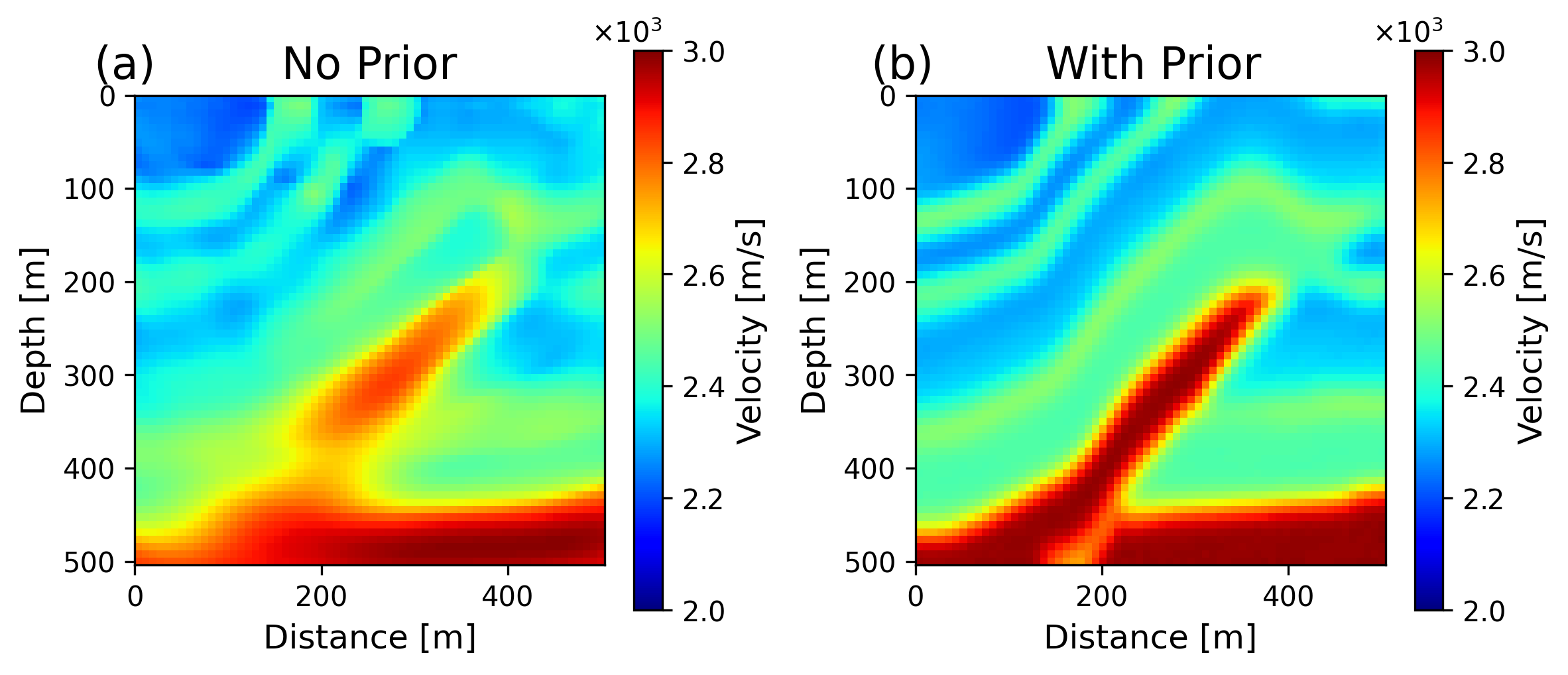}
    \caption{\textbf{a} Inversion result using conventional FWI (no prior) and \textbf{b} inversion result using our proposed diffusion-prior FWI with regularization term (with prior).}
    \label{fig:inverse}
\end{figure}

We use the Adam optimizer \citep{kingma2014adam} with a learning rate of $1\times10^{-2}$ for both conventional FWI and diffusion-prior FWI and run 100 iterations in each case. Figure~\ref{fig:inverse} compares the final inversion results from conventional and diffusion-prior approaches. The conventional FWI result demonstrates limited accuracy, particularly in deeper regions and at lateral boundaries, indicating convergence to a local minimum in the absence of prior information.

\begin{figure*}[!htp]
    \centering
    \includegraphics[width=1.0\textwidth]{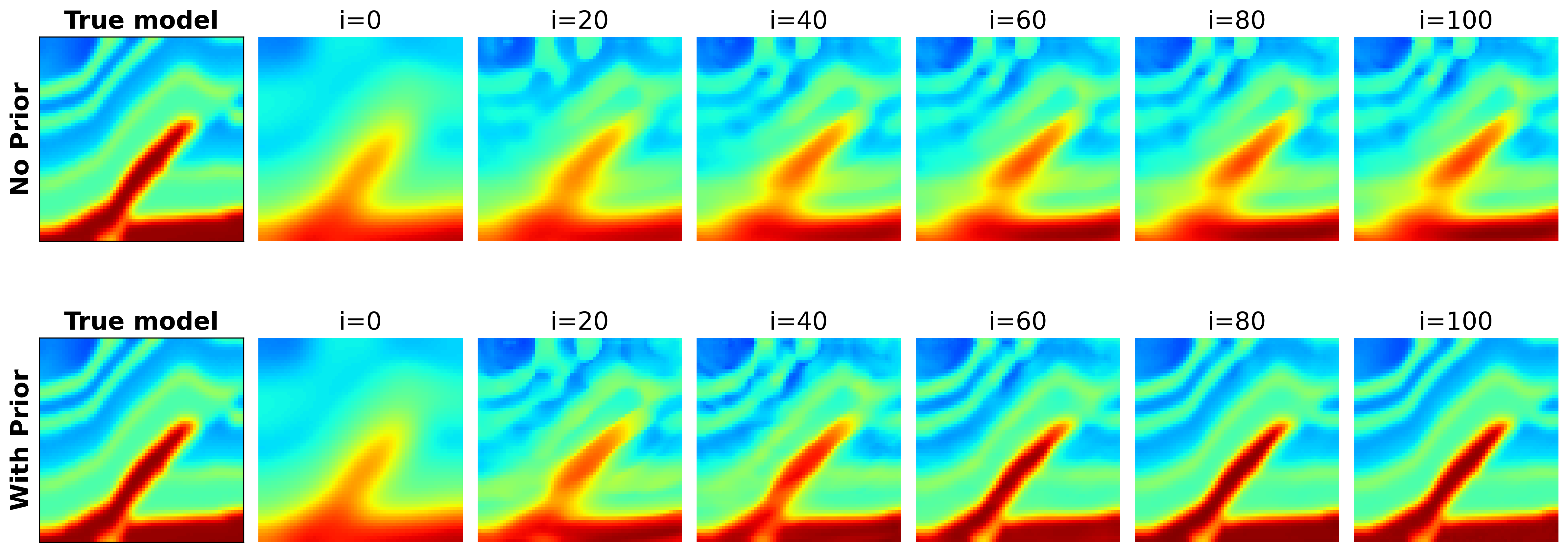}
    \caption{Intermediate inversion results across selected iterations, comparing conventional FWI (no prior) and diffusion-prior FWI (with prior), with the same true model (top left) shown for reference.}
    \label{fig:intermediate}
\end{figure*}

Figure~\ref{fig:intermediate} shows selected intermediate results from both methods throughout the optimization process starting from the same initial velocity model. The diffusion-prior FWI maintains consistency with the data manifold, resulting in smoother transitions and more geologically plausible structures. In contrast, conventional FWI produces artifacts and unstable updates due to its lack of regularization. The incorporation of the learned prior in diffusion-prior FWI guides the inversion toward realistic subsurface structures while preserving fidelity to the observed seismic data.

\begin{figure}[!htp]
    \centering
    \includegraphics[width=0.5\textwidth]{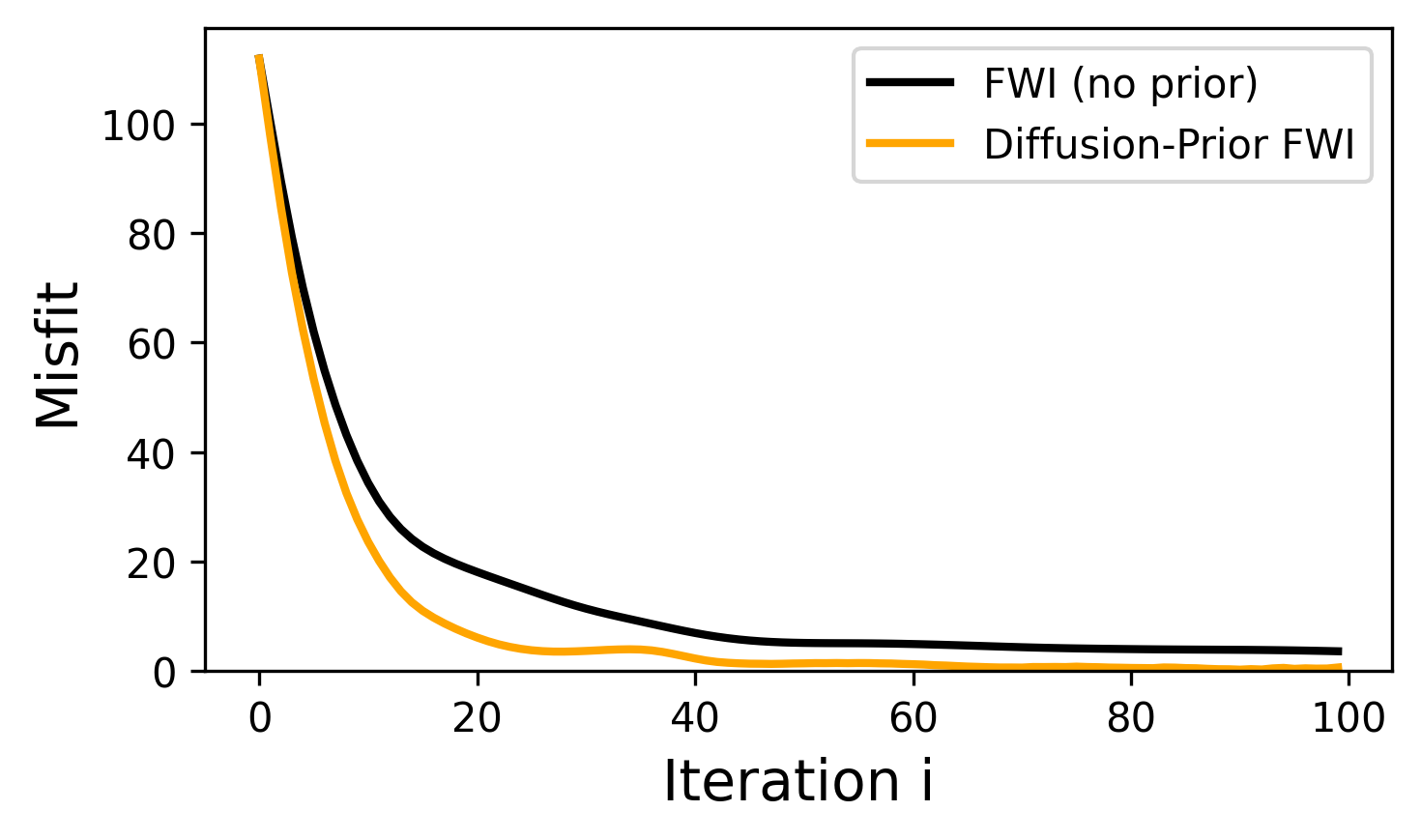}
    \caption{Data misfit history measured by the $\ell_2$ norm for conventional FWI (without prior) and diffusion-prior FWI (with prior). The diffusion-prior FWI demonstrates faster and more stable convergence, achieving a lower final data misfit.}
    \label{fig:loss}
\end{figure}

Figure~\ref{fig:loss} presents the data misfit history, quantified by the $\ell_2$ norm, for both conventional FWI and diffusion-prior FWI. The diffusion-prior FWI achieves a faster convergence rate and consistently lower misfit values throughout the inversion process. The final misfit is also reduced compared to conventional FWI, indicating the effectiveness of the learned prior in guiding the optimization toward a more accurate solution with a more realistic inversion result.

\begin{figure*}
    \centering
    \includegraphics[width=0.75\textwidth]{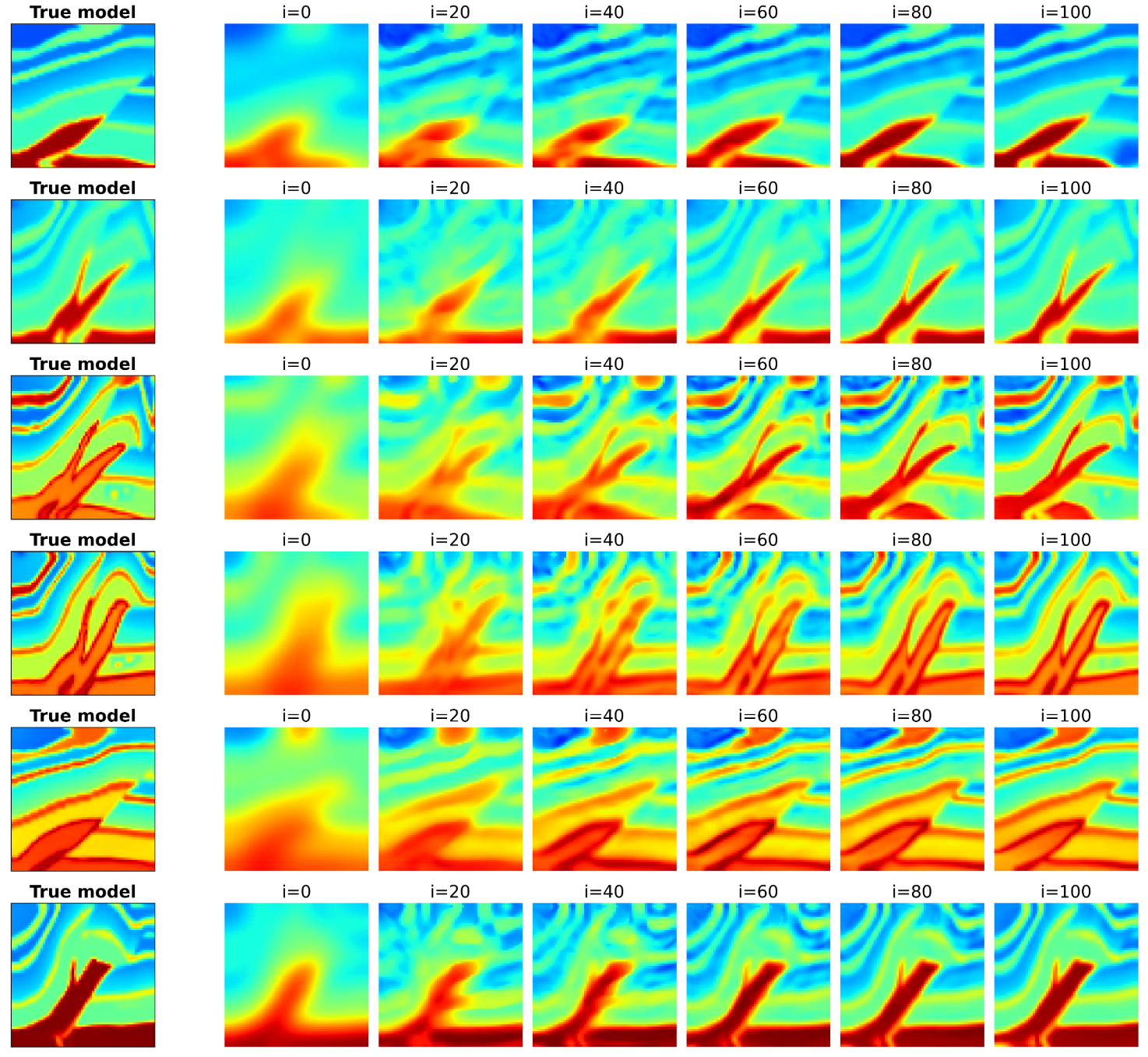}
    \caption{Additional tests using different initial velocity models (left), evaluated under the same optimization configuration described earlier. The figures show the updated velocity fields after 100 iterations of the diffusion-prior FWI.}
    \label{fig:tests}
\end{figure*}

Figure~\ref{fig:tests} presents additional experiments conducted using the proposed diffusion-prior FWI framework, trained on the Overthrust dataset. Despite the variation in initial velocity models and true models, the inversion results consistently converge toward geologically plausible structures and align well with the corresponding true models. These results show the robustness and generalization ability of the learned diffusion prior across different initializations.

\section{Generalization Capability of the Pretrained Diffusion regularizer}

While the diffusion FWI regularizer shows strong performance on synthetic benchmarks, it is important to evaluate its generalization to unseen datasets. To this end, we tested the mentioned pretrained diffusion prior, originally trained on the Overthrust dataset, on a section of the Marmousi2 dataset \citep{martin_marmousi2_nodate}. Despite the domain shift, the pretrained model remained adaptable. Although the Overthrust-trained prior had implicitly learned features such as sharp subsurface contrasts, which may not exactly match the Marmousi2 geological patterns, it still improved structural recovery. The regularization introduced by the diffusion prior produced more geologically plausible velocity models than conventional FWI without regularization, particularly in deeper regions where data fit gradients are weak.

\subsection{Training a Generalized Diffusion Prior for Broader Inversion Tasks}

To further improve the generalization capability of the diffusion prior, we trained a new DDPM model on a more diverse and synthetic set of subsurface structures. Specifically, we developed a Gaussian process-based random image generator to synthesize velocity models with continuous geological features and varying contrasts. These random fields were governed by a covariance matrix $Q$, which controls the spatial correlation of subsurface properties, allowing for the creation of complex geological variations. We applied geometric data augmentation techniques to introduce some faulted layers, thereby expanding the diversity of structural patterns.

\begin{figure}[!htp]
    \centering
    \includegraphics[width=0.4\textwidth]{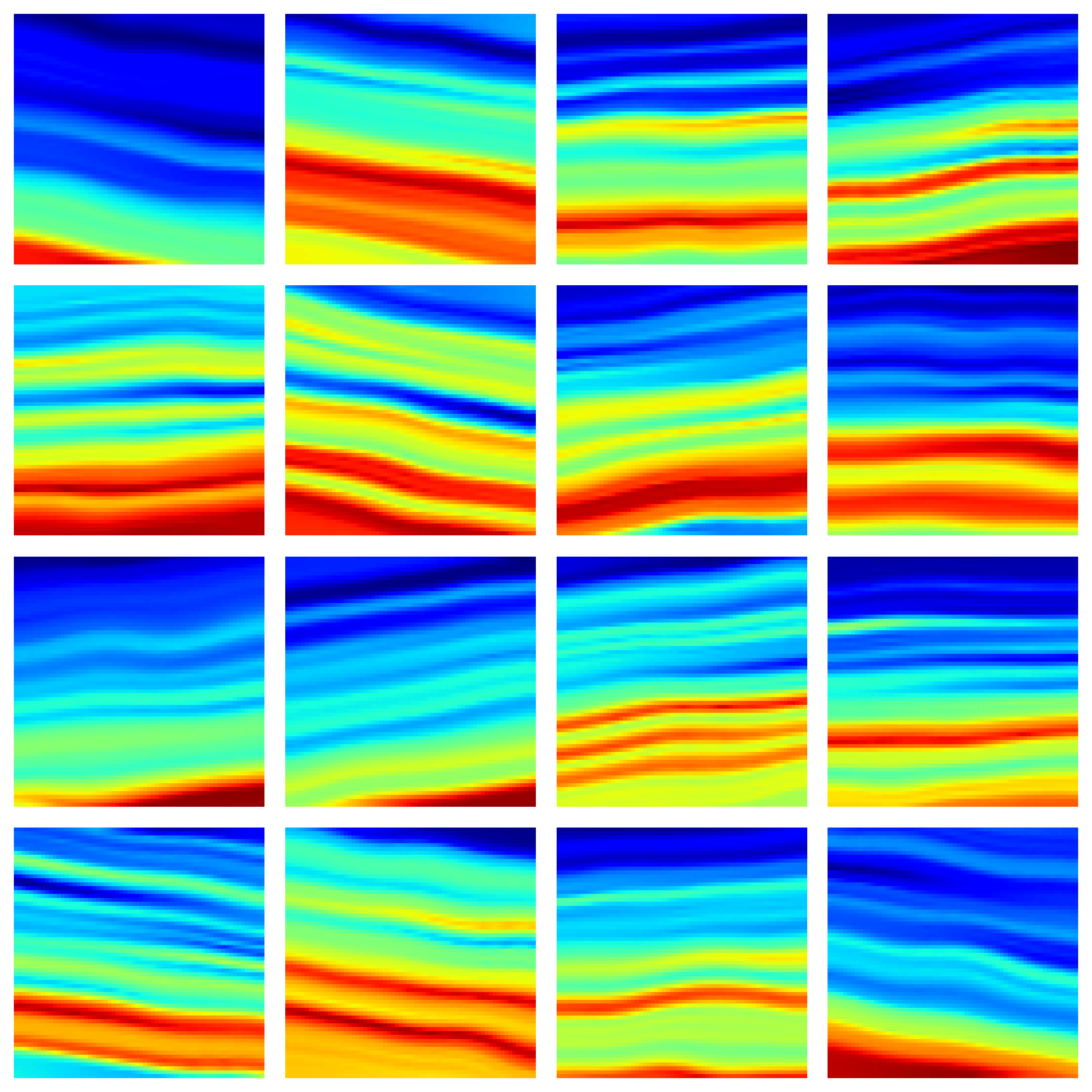}
    \caption{Samples generated by retraining the DDPM on 100,000 Gaussian process-based images, demonstrating its ability to capture a wider variety of plausible geological structures.}
    \label{fig:gaussian_prior}
\end{figure}

In total, we generated 100,000 synthetic subsurface images using this approach. The new DDPM trained on these Gaussian process-based images acts as a more flexible prior, applicable to a broader range of inversion tasks.

When applied to the retrained diffusion prior to the Marmousi2 dataset, this retrained diffusion prior significantly improved inversion performance, compared to the Overthrust-trained prior. The generalized model provided better recovery of structural contrasts and enhanced deeper sections of the model compared to the conventional FWI without regularization. This improvement is particularly notable in areas where the wavefield sensitivity is low and traditional adjoint gradient-based updates tend to underperform.

\begin{figure*}[!htp]
    \centering
    \includegraphics[width=1\textwidth]{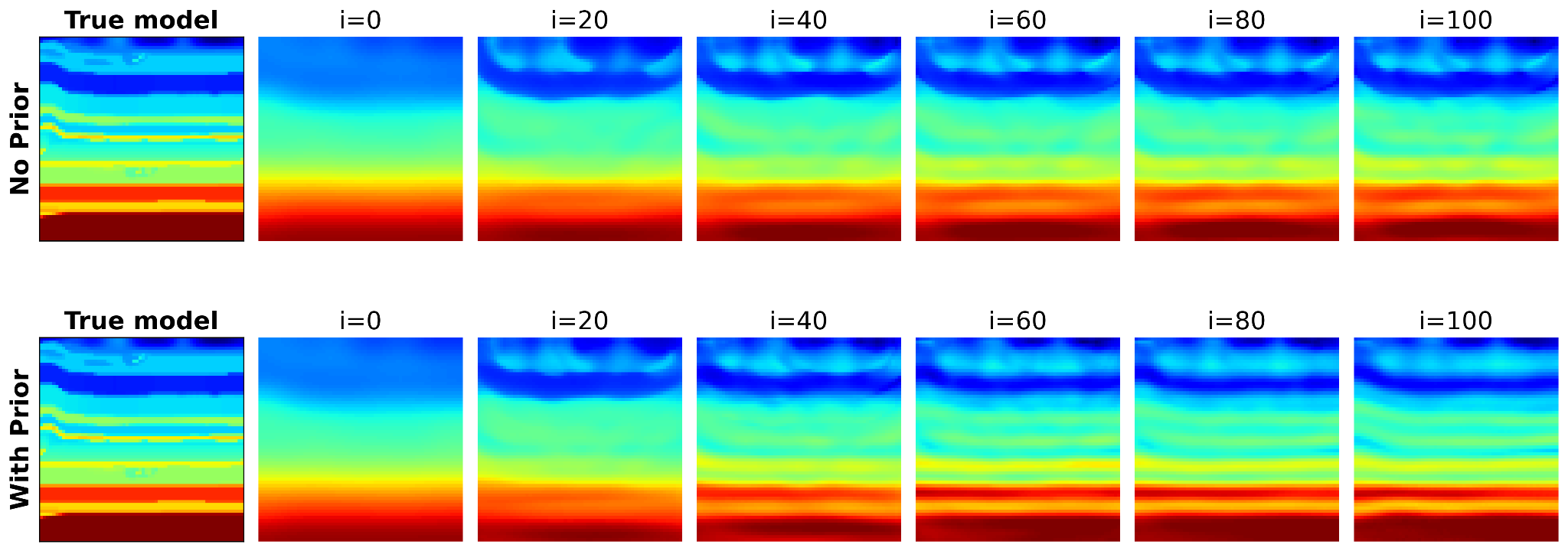}
    \caption{Inversion result on Marmousi2 dataset using the DDPM retrained on Gaussian process-generated images. The model exhibits improved structural detail and contrast, particularly in deeper regions.}
    \label{fig:gaussian_prior_fwi_result}
\end{figure*}

\section{Discussion}

The proposed method demonstrates strong potential for leveraging pretrained DDPMs and other score-matching-based diffusion models as generative priors. Benefiting from the rapid growth of the generative diffusion model community, a wide range of high-quality pretrained models are now available for various image generation tasks. In contrast to traditional diffusion-based approaches, our framework does not require solving the full reverse diffusion process. Instead, it selectively utilizes intermediate diffusion time steps and performs score-matching using a fixed, pretrained DDPM denoiser to iteratively update the velocity model through gradient-based optimization.

This gradient-driven score-matching formulation enables the effective incorporation of prior knowledge from the data manifold while remaining compatible with wave propagation solvers and other nonlinear physical models that demand numerical stability. In the context of seismic imaging and FWI, our method avoids the emergence of false reflectors and suppresses non-physical scattering artifacts, common issues arising when noise is directly injected into velocity models. Such artifacts can compromise stability and degrade inversion quality, especially under finite-difference simulation frameworks. By preserving physical consistency and leveraging learned priors, the proposed approach supports stable, high-fidelity inversion across iterations.

\subsection{Differentiability of the Neural Network}

As outlined in the Theory section, the proposed method requires differentiating through a pretrained neural network—in this case, a U-Net architecture trained to approximate the score function of the diffusion model. This differentiability is essential to compute the gradient of the score-matching term with respect to the velocity model. Like many automatic differentiation-based optimization methods, the primary computational bottleneck lies in the memory demands associated with backpropagation through deep networks.

In our experiments, we utilized an 8GB NVIDIA GPU, which enabled processing of a batch of 32 images with gradient tracking through the network. This setup was sufficient for our current study. However, scaling the method to higher-resolution velocity models, deeper neural networks, or more expressive generative models (e.g., stable diffusion or latent diffusion architectures) would significantly increase memory requirements. Future extensions may require memory-efficient differentiation strategies or distributed computing resources to maintain feasibility in large-scale geophysical applications.

\subsection{Flexibility Compared to GAN-Based Priors and the Role of Prior Weighting}

Compared to methods that use GANs or other generative models with strict priors, where solutions are forced to lie exactly on the generator manifold, often making the optimization problem more highly non-linear, our diffusion-based approach provides greater flexibility. The use of a diffusion prior allows for a softer regularization mechanism, where the influence of the prior can be continuously modulated via a weighting parameter. This enables the inversion to explore intermediate solutions that balance data misfit with prior consistency, offering a continuum between conventional FWI results and strongly regularized outputs aligned with the learned manifold.

This balance is particularly advantageous in practical applications, where strict adherence to the learned prior may suppress data-consistent features, while unregularized inversions may overfit noise or produce geologically implausible artifacts. However, the effectiveness of this balance is sensitive to the choice of the regularization weight. A suboptimal weight may lead to either an under-regularized result with unstable wave propagation or an over-regularized output that suppresses genuine subsurface features. Thus, careful calibration of this hyperparameter is essential, and adaptive strategies may be considered in future work to dynamically adjust the influence of the prior during inversion.

\subsection{Importance of the Prior and Training Data in Bayesian Inversion}

In any Bayesian inversion framework, the specification of the prior distribution plays a critical role in shaping the posterior solution. The prior encapsulates our assumptions and knowledge about the physical plausibility of the model space. In the context of our method, the learned diffusion prior serves as a probabilistic model of the data manifold, encoding spatial patterns and geological features derived from training data. This offers a powerful means to regularize the inversion and improve stability, especially in ill-posed regimes or under sparse or noisy observations.

However, the efficacy of this regularization is inherently dependent on the representativeness of the training dataset. If the training images do not adequately capture the variability or complexity of the true subsurface structures, the prior may bias the inversion toward geologically unrealistic solutions. Thus, the choice and curation of training data are of paramount importance in ensuring that the learned prior both enhances image quality and retains geophysical relevance. While this offers a strong inductive bias to guide inversion, it also restricts the solution space, potentially limiting the flexibility of the posterior in accommodating novel or out-of-distribution features.

\subsection{Uncertainty and interpretability in Learned Priors}
An important aspect to consider when integrating learned priors, such as those derived from DDPMs, into geophysical inversion workflows is the interpretability of the resulting models and the quantification of uncertainty. While conventional FWI provides deterministic outputs, generative diffusion models have the potential to provide a probabilistic inversion in a Bayesian framework.

However, in the current implementation, the diffusion FWI regularizer is used in a gradient-based update framework without explicitly sampling the posterior or capturing the spread of uncertainty. This means that while the learned prior serves similar to regularization in the inversion toward geologically plausible structures, it does not yet provide uncertainty estimates in a principled way. Future extensions could explore combining the framework with posterior sampling techniques, such as Langevin dynamics or Hamiltonian Monte Carlo under a score-based prior to samples from the full conditional posterior \citep{zhang2024improving}, allowing the characterization of model uncertainty and ambiguity.

Moreover, interpretability remains a challenge. The influence of the prior is learned from data, and while this often enhances realism and stability, it may be less transparent than traditional regularization methods (e.g., Tikhonov or total variation). Developing tools to visualize and quantify how the prior shapes the inversion and to assess when it aligns or conflicts with the observed data will be crucial in increasing practitioner trust and understanding of such hybrid inversion approaches.

\section{Conclusion}

We propose a novel deep learning approach to integrate pretrained diffusion models into FWI as a simple regularization term, operating directly in the clear image space without requiring reverse diffusion sampling and without operations in noisy images.

Our method introduces a generative diffusion FWI regularizer using the score rematching gradient that leverages learned data priors while preserving data fit to seismic observations through physics-based misfit terms using a standard FWI gradient. Numerical experiments demonstrate improved convergence and stability compared to GAN, as well as better inversion quality over conventional FWI.

The proposed framework offers a simple, flexible, and effective way to integrate generative priors into traditional FWI workflows, with minimum modification to the current FWI workflow. Our approach opens up new avenues for future work, particularly the extension to elastic FWI and uncertainty quantification.

\bibliographystyle{abbrvnat}
\bibliography{references}

\end{document}